\pdfoutput=1

\documentclass[twocolumn,prl,showpacs,amsmath,amssymb,superscriptaddress]{revtex4}

\usepackage{bm}
\usepackage{graphicx}
\usepackage{amssymb}
\usepackage{xcolor}

\begin{document}

\title{Zero-Field Quantum Critical Point in Ce$_{0.91}$Yb$_{0.09}$CoIn$_5$}


\author{Y. P. Singh\footnote{These authors have contributed equally to this work.} }
\altaffiliation[Present Address: ]{Department of Mechanical Engineering, The University of Akron, Akron, Ohio, 44325, USA}
\affiliation{Department of Physics, Kent State University, Kent, Ohio, 44242, USA}

\author{R. B. Adhikari$^*$ } 
\affiliation{Department of Physics, Kent State University, Kent, Ohio, 44242, USA}

\author{D. J. Haney}  
\affiliation{Department of Physics, Kent State University, Kent, Ohio, 44242, USA}

\author{B. D. White}
\affiliation{Center for Advanced Nanoscience, University of California, San Diego, La Jolla, California 92093, USA}
\affiliation{Materials Science and Engineering Program, University of California, San Diego, La Jolla, California 92093, USA}
\affiliation{Department of Physics, University of California at San Diego, La Jolla, CA 92903, USA}

\author{M. B. Maple}
\affiliation{Center for Advanced Nanoscience, University of California, San Diego, La Jolla, California 92093, USA}
\affiliation{Materials Science and Engineering Program, University of California, San Diego, La Jolla, California 92093, USA}
\affiliation{Department of Physics, University of California at San Diego, La Jolla, CA 92903, USA}

\author{M. Dzero}
\affiliation{Department of Physics, Kent State University, Kent, Ohio, 44242, USA}

\author{C. C. Almasan}
\affiliation{Department of Physics, Kent State University, Kent, Ohio, 44242, USA}

\date{\today}

\begin{abstract}
We present results of specific heat, electrical resistance, and magnetoresistivity measurements on single crystals of the heavy-fermion superconducting alloy Ce$_{0.91}$Yb$_{0.09}$CoIn$_5$. Non-Fermi liquid to Fermi liquid crossovers are clearly observed in the temperature dependence of the Sommerfeld coefficient $\gamma$ and resistivity data. Furthermore, we show that the Yb-doped sample with $x=0.09$ exhibits universality due to an underlying quantum phase transition without an applied magnetic field by utilizing the scaling analysis of $\gamma$.  Fitting of the heat capacity and resistivity data based on existing theoretical models indicates that the zero-field quantum critical point is of antiferromagnetic origin. Finally, we found that at zero magnetic field the system undergoes a third-order phase transition at the temperature $T_{c3}\approx 7$ K. 
\end{abstract}
\maketitle

\paragraph{Introduction.} The study of quantum critical behavior of modern materials continues to be a central topic in condensed-matter physics since quantum phase transitions (QPTs) at a quantum critical point (QCP) can drive a system away from its normal metallic behavior, resulting in distinctly different physical properties in the vicinity of QCP \cite{Coleman2005,VojtaReview2007}. 
In unconventional superconductors (SC), such as heavy fermions (HF) materials, cuprates, and pnictides, the presence of competing interactions, due to the proximity of antiferromagnetism and superconductivity, can give rise to zero-point critical fluctuations and to a QPT from a magnetically ordered to a disordered phase. This raises the possibility that the unconventional superconducting pairing in these systems is mediated by antiferromagnetic (AFM) spin fluctuations. In addition, many of these systems exhibit deviation from their Fermi-liquid  properties in the presence of a QCP. Hence these materials offer great potential to study and understand the nature of unconventional superconductivity. 

Ce$_{1-x}$Yb$_x$CoIn$_5$ is an intriguing HF system which has attracted much attention because many of the properties observed in this material do not conform to those of similar HF superconductors \cite{Shu2011,White2012,Xu2016}. The parent compound CeCoIn$_5$ is an example of a metal in which the system's proximity to a QPT between the paramagnetic and AFM states is controlled by thermodynamic variables such as magnetic field ($H$) or pressure, with the antiferromagnetic ground state superseded by superconductivity \cite{Bianchi2003,Sarrao2007,Zaum2011,Hu2012}. Substitution of Ce by Yb results in a gradual suppression of the magnetic-field-driven QCP (H$_{QCP}$) to zero in the vicinity of the Yb doping level $x = 0.07$ \cite{Hu2013}, suggesting that there exists a critical concentration $x_c$ at which the low-$T$ properties of Ce$_{1-x_c}$Yb$_{x_c}$CoIn$_5$ are of a quantum critical metal. In addition, recent reports reveal a significant modification of the Fermi surface of Ce$_{1-x}$Yb$_x$CoIn$_5$ with Yb substitution \cite{Polyakov2012} and its doping-dependent change from a nodal to a nodeless gap \cite{Kim2015}. All these studies have led the way to proposals of new, exciting possibilities such as composite pairing mechanism and topological superconductivity~\cite{Erten2015} and two different Fermi surfaces contributing to charge transport~\cite{Singh2015}. However, an extremely important question is  pertaining to the role played by quantum critical fluctuations in determining the symmetry of the superconducting gap and system's thermodynamic and transport properties. 

Here, we present results of thermodynamic and transport studies of the superconducting alloy Ce$_{0.91}$Yb$_{0.09}$CoIn$_5$ tailored to address the above question in the Ce$_{1-x}$Yb$_{x}$CoIn$_{5}$ heavy-fermion family. Our study reveals a clear correlation between critical spin fluctuations and unconventional superconductivity. Specifically, the presence of a crossover from a Fermi liquid (FL) to a non-Fermi liquid (NFL) regime along with the scaling analysis of the Sommerfeld coefficient demonstrate that the normal state of Ce$_{0.91}$Yb$_{0.09}$CoIn$_5$ is quantum critical. Furthermore, our analysis reveals that the underlying quantum phase transition is of magnetic nature separating antiferromagnetic and paramagnetic phases. These present results along with the results from Ref. \cite{Kim2015} indicate that the nodal gap structure and unconventional superconductivity in Ce$_{1-x}$Yb$_{x}$CoIn$_{5}$ for Yb doping $x$ smaller than the critical value $x_c$ is due to the presence of AFM critical spin fluctuations near a QCP, while for $x>x_c$ the critical fluctuations are fully suppressed and this system displays conventional SC.  Our present results also show that quantum criticality emerges for Yb concentrations for which Yb is in a magnetic valence configuration that seems to play a crucial role in preserving the long-range order of the diluted Ce lattice and stabilizing the unconventional superconducting state in this fascinating material. Finally, we also found that at $H=0$ the system undergoes a third-order phase transition at the temperature $T_{c3}\approx 7$ K. 

\begin{figure}
\centering
\includegraphics[width=1.0\linewidth]{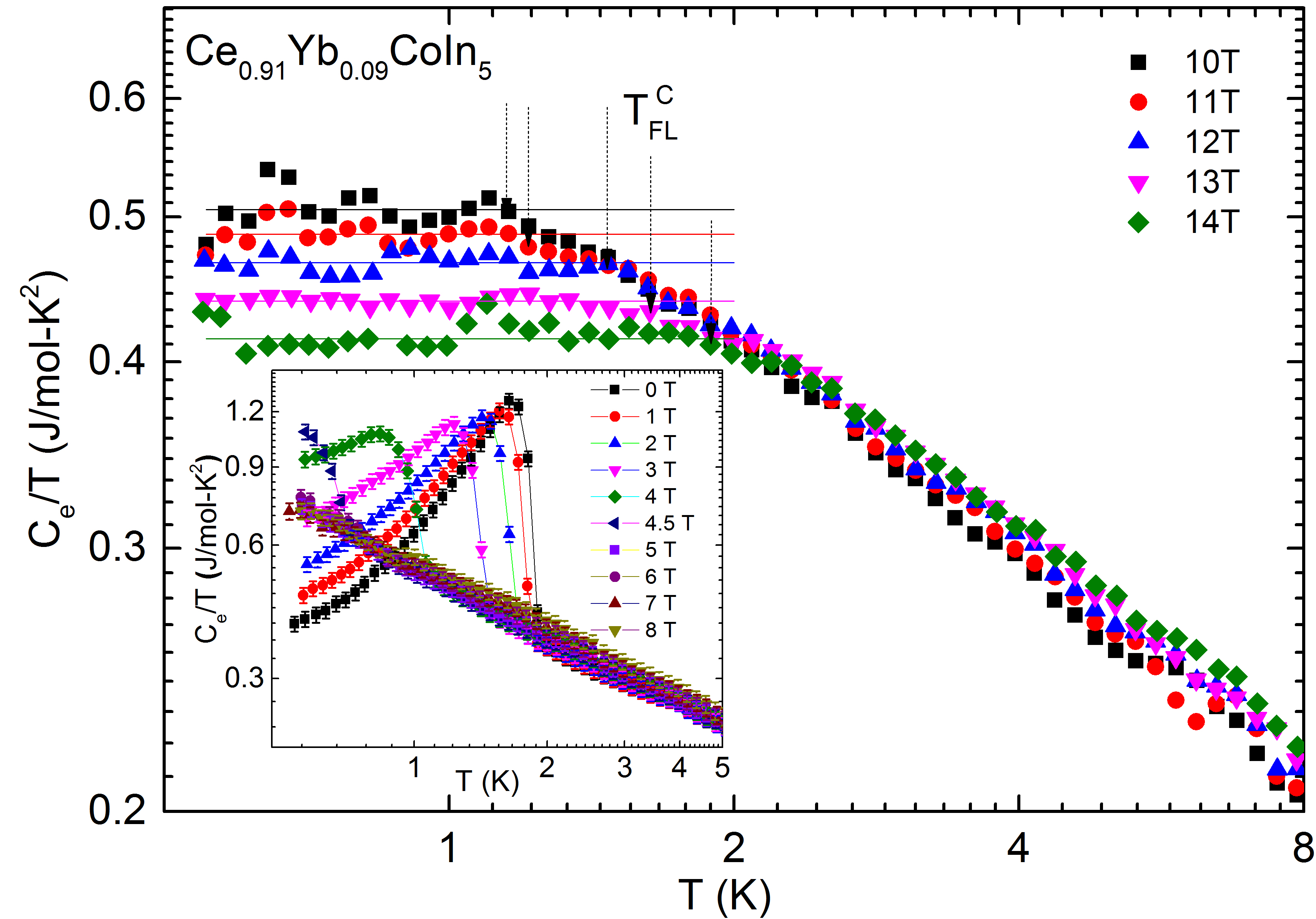}
\caption{\small(Color online) Log-log plot of Sommerfeld coefficient $\gamma \equiv C_e(T)/T$ ($C_e$ is the electronic specific heat) vs. temperature $T$ of Ce$_{0.91}$Yb$_{0.09}$CoIn$_{5}$ measured with applied magnetic field $H \parallel c$-axis over the temperature range $0.50 \leq T \leq 10$ K and  $10 \leq H \leq 14$ T. The horizontal lines are fits of the data. The temperature $T_{\textrm{FL}}^C$, marked by the arrows, is the crossover temperature from the Fermi liquid ($T$ independent $\gamma$) to the non-Fermi liquid  ($\gamma$ has power-law-in-$T$ dependence) regimes. Inset: Log-log plot of $C_e/T$ vs $T$, measured in $0 \leq H \leq 8$ T.}
\label{fig:HighHC09}
\end{figure} 

\paragraph{Experimental details.} Single crystals of Ce$_{0.91}$Yb$_{0.09}$CoIn$_5$ were grown using an {indium self-flux method \cite{Moshopoulou2001}}. They were etched in concentrated HCl to remove the indium left on the surface during the growth process and were then rinsed thoroughly in ethanol.
The crystal structure and unit cell volume were determined from X-ray powder diffraction measurements, while the actual Yb composition $x$ for the samples studied here was determined according to the method developed by Jang et al.~\cite{Jang2014}. We note that we will refer to nominal Yb concentration $x_n$ when we mention previous results that report only this value. The single crystals studied have a typical size of $2.1 \times 1.0 \times 0.16$ mm$^3$, with the $c$-axis along the shortest dimension of the crystals. 

Heat capacity measurements were performed under magnetic fields up to  14 T, applied parallel to the $c$ axis (${H \parallel c}$) of the crystals and for temperatures as low as 0.50 K. The data was obtained in a Quantum Designs Physical Property Measurement System using semi-adiabatic calorimetry and utilizing the heat pulse technique. 

Four leads were attached to the single crystals, with the current $I \parallel a$-axis, using a silver-based conductive epoxy. 
We performed in-plane electrical resistance ($R$) measurements between 0.50 and 300 K and transverse magnetoresistivity (MR) ${\bigtriangleup\rho \equiv [\rho(H)-\rho(H=0)]/\rho(H=0)}$ measurements as a function of temperature between 2 and 300 K and transverse magnetic field (${H \perp I}$) up to 14 T.
 
\paragraph{Results and discussion.} In Fig.~\ref{fig:HighHC09} we show the temperature dependence of the Sommerfeld coefficient $\gamma \equiv C_e(T)/T$ of Ce$_{0.91}$Yb$_{0.09}$CoIn$_5$ measured in $10 \leq H \leq 14$ T and at low temperatures ($0.5 \leq T \leq 8$ K). We obtained the electronic specific heat $C_e$  after subtracting a large Schottky anomaly tail due to the quadrupolar and magnetic spin splitting of Co and In nuclei \cite{Movshovich2001}. All the data shown in this figure represent normal-state results since the superconducting transition temperature is gradually suppressed with increasing magnetic field and approaches 0.5 K (the lowest temperature of the heat capacity measurements performed here) for a field of 4.5 T (see inset to Fig.~\ref{fig:HighHC09}). Notice that all the data shown in the main panel of Fig.~\ref{fig:HighHC09} display a crossover between a constant $\gamma$ at low temperatures and a power-law $T$ dependent $\gamma$ at higher $T$, with $\gamma \approx 0.52/T^{0.48}$, supported by the data analysis shown in the left inset of Fig.~\ref{fig:Scaling09}. The former behavior is typical of the FL state, while the latter reflects a non-Fermi liquid state. We define as $T_{\textrm{FL}}^\textrm{C}$ (arrows in the main panel of Fig.~\ref{fig:HighHC09}) the temperature at which $\gamma(T)$ deviates from the horizontal line, i.e. where it changes from the FL region to the NFL (power-law-in-$T$ dependence) region. 

The presence of critical fluctuations as well as the FL to NFL crossover observed in both heat capacity and resistivity (see Supplementary Materials for the analysis of resistivity and heat capacity data) suggest the presence of a second-order QPT at a QCP. Quantum phase transitions are different from conventional thermodynamic transitions in that the correlations of the incipient order parameter fluctuate on a characteristic energy scale 
$E_0\gg k_BT_c$, where $T_c$ is the critical temperature, which in our system is vanishingly small. This energy scale $E_0$ becomes also vanishingly small as the host system is tuned to QCP  \cite{VojtaReview2007} by varying thermodynamic variables such as magnetic field and pressure or by changing the chemical composition, and temperature remains the only energy scale which controls the physics at low temperatures. As a result, the system's thermodynamic properties are dominated by the continuum of thermally excited quantum critical fluctuations at low enough temperatures, $k_BT\ll E_0$. Consequently, the specific heat and magnetic susceptibility exhibit anomalous power-law temperature dependences, which can be accounted for by exponents whose values are determined by the nature of the order parameter fluctuations and the relative strength of the interactions between the quantum critical fluctuations. Perhaps the most recent example of this has been provided by W\"{o}lfle and Abrahams who have argued  that an interplay between the non-gaussian quantum critical fluctuations and itinerant fermionic quasiparticles leads to the anomalous temperature dependence of the Sommerfeld coefficient $\gamma\propto 1/T^{0.25}$ in the quantum critical metal YbRu$_2$Si$_2$ \cite{Elihu2011,JoergReview2016}.

\begin{figure}
\centering
\includegraphics[width=0.8\linewidth]{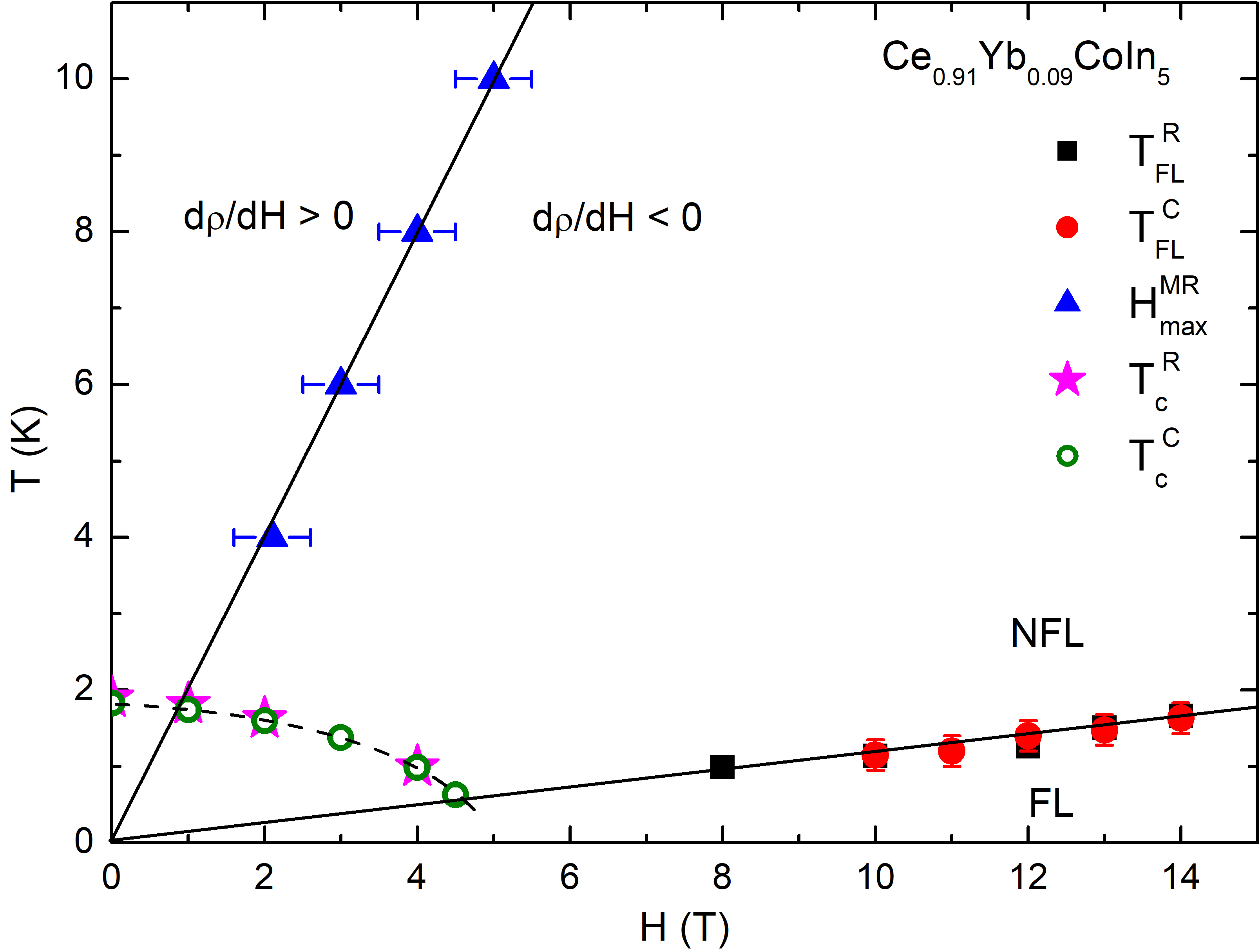}
\caption{\small(Color online) Temperature - magnetic field $T-H$ phase diagram of Ce$_{0.91}$Yb$_{0.09}$CoIn$_{5}$ with ${H \parallel c}$-axis. The area under dotted line represents the superconducting region. The straight lines are linear fits of the data extrapolated to zero temperature.  $H_{max}^{MR}$ is the peak in the $H$ dependence of magnetoresistivity MR (see Supplementary Materials for details).  $T_{FL}^R$ and $T_{FL}^C$ are the temperature at which the data cross from a Fermi liquid to a non-Fermi liquid regime, measured by electrical resistance and heat capacity, respectively.  Likewise, $T_{c}^R$ and $T_{c}^{C}$ is the superconducting transition temperature measured by electrical resistance and heat capacity, respectively.}
\label{fig:PhaseDiagram09}
\end{figure}

\begin{figure}
\centering
\includegraphics[width=0.8\linewidth]{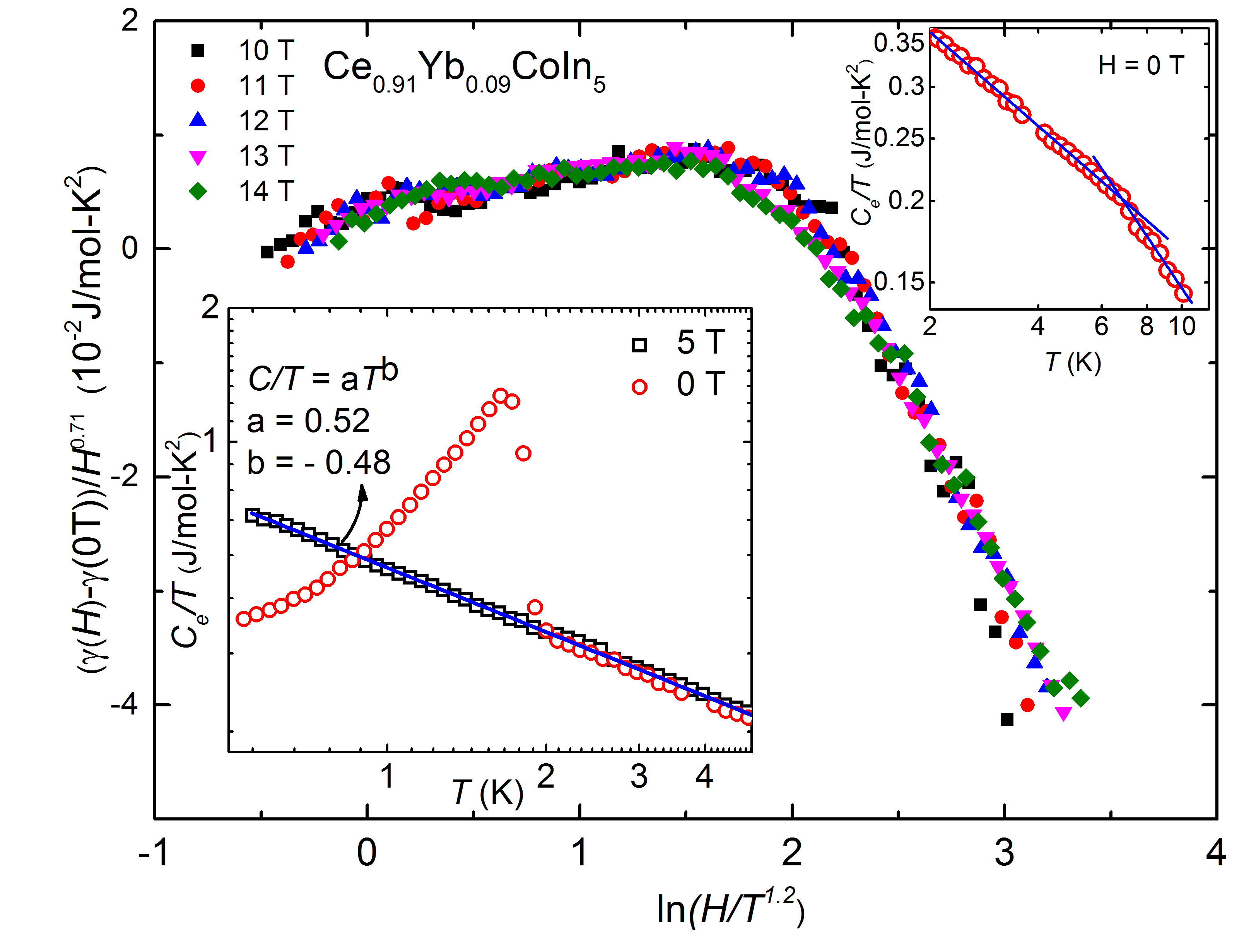}
\caption{\small(Color online) Scaling of $\gamma$ according to the function $\gamma(H)-\gamma(H_{\textrm{QCP}})\sim (H-H_{\textrm{QCP}})^{0.71}f\left((H-H_{\textrm{QCP}})/T^{1.2}\right)$ with $H_{QCP}$ = 0 T. We obtained the best scaling shown on the main panel with a power-law dependence of $\gamma(0)$ at temperatures $T\leq 5$ K. Bottom inset: Log-Log plot of 0 T and 5 T data and their normal state fit with $C_e/T$=$0.52/T^{0.48}$. Top inset: Log-log plot of $C_e/T$ vs $T$ data measured in $H=0$ to show third-order phase transition.} 
\label{fig:Scaling09} 
\end{figure}

In order to verify that the concentration $x_c=0.09$ corresponds to a critical value at which $E_0(x=x_c)\to 0$, we generated the $H$-$T$ phase diagram shown in Fig.~\ref{fig:PhaseDiagram09}. The FL to NFL crossovers are extracted from heat capacity and resistance measurements as $T_{\textrm{FL}}^\textrm{C}(H)$ (black squares) and $T_{\textrm{FL}}^R(H)$ (red circles), respectively, as discussed above and in the Supplementary Materials section. The $T_c(H)$ boundary separating metallic and superconducting phases is obtained from the resistance and heat capacity measurements. Notice that the FL to NFL crossovers obtained from the two measurements are in excellent agreement. Linear extrapolations of these crossovers to zero temperature, under the superconducting dome, give the value of $H_{\textrm{QCP}}$. Figure~\ref{fig:PhaseDiagram09} clearly shows that these linear extrapolations give $H_{\textrm{QCP}}=0$, indicating that 9$\%$ Yb is the critical doping, i.e., $x_c=0.09$ for the Ce$_{1-x}$Yb$_{x}$CoIn$_{5}$ system. 
Additional evidence comes from our analysis of magnetoresistivity (MR) data measured in different applied magnetic fields (see Supplementary Materials section). Specifically, a linear extrapolation of the peak in MR at $H_{max}^{MR}$ (blue triangles in Fig.~\ref{fig:PhaseDiagram09}) to zero temperature again results in $H_{\textrm{QCP}} = 0$ T. It is also worth mentioning here that the Ce$_{1-x}$Yb$_{x}$CoIn$_{5}$ system exhibits a crossover at $x = 0.09$ from a Kondo lattice ($x<0.09$) to a single-ion Kondo regime ($x>0.09$)  \cite{Singh2014}. In other words, the Ce ions in the latter case do not form a lattice, hence, the possibility of a magnetically ordered phase can be ruled out for Yb doping $x > 0.09$. 

Finally, in order to confirm that the anomalous temperature dependence of the Sommerfeld coefficient $\gamma(H,T)$ 
is governed by quantum critical fluctuations, we show that $\gamma(H,T)$ is governed by the critical free energy density $f_\textrm{cr}=a_0r^{\nu(d+z)}f_0(T/r^{\nu z})=a_0T^{(d+z)/z}\tilde{f}_0(r/T^{1/\nu z})$, where
$a_0$ is a constant, $f_0$ and $\tilde{f}_0$ are scaling functions, $r\propto (H-H_{\textrm{QCP}})$, $d$ is the dimensionality of the system, $z$ is the dynamical critical exponent, and $\nu$ is the correlation length exponent. By comparing the Sommerfeld coefficient $\gamma_{\textrm{cr}}=(\partial^2 f_{\textrm{cr}}/\partial T^2)$ 
with the experimentally measured one at $H=0$ (inset to Fig.~\ref{fig:Scaling09}), we find
$d=0.52z$. Furthermore, at finite magnetic fields, we were able to scale the data by choosing $\nu(d-z)=0.71$ and $1/\nu z=1.2$ (see Fig.~\ref{fig:Scaling09}). We note that the $1/\sqrt{T}$ divergence of the Sommerfeld coefficient is fully consistent with the universal restriction $1/\nu z<3/2$ \cite{Tsvelick1993}. Lastly, we point out that the critical exponent $\alpha=1+d/z$ for $\gamma(T,H=0)\sim T^{-\alpha}$ is governed by the universality class of ordinary percolation. 

The result of the scaling is shown in Fig.~\ref{fig:Scaling09}. We obtained a very good scaling with the normal-state zero-field $\gamma(T,0)=0.52/T^{b}$, with $b=0.48$.  The scaling of $\gamma$ covers both the FL range (present at low $T$) and the NFL range (present at higher $T$), with a small amount of scatter near  the FL to NFL  crossover temperature. We attribute this to the lack of a sharp crossover for magnetic fields of 13 T, and 14 T, as seen in Fig.~\ref{fig:HighHC09}. A peculiar feature of our scaling plot is the existence of a hump for $\log(H/T^{1.2})\in[1.5,2]$ followed by a decrease at higher temperatures. This is a result of the power-law divergence of the Sommerfeld coefficient at the quantum critical field, which in this case is $H_{QCP}=0$.
This feature should be clearly absent if the Sommerfeld coefficient has a logarithmic divergence with temperature, as is the case for CeCoIn$_5$ (see Fig. 1(b) of Ref. \cite{Bianchi2003}).

We note that although both Ce$_{0.91}$Yb$_{0.09}$CoIn$_{5}$ and YbRu$_2$Si$_2$ have antiferromagnetic QCPs, the value of the exponent $b=0.48$ in Ce$_{0.91}$Yb$_{0.09}$CoIn$_{5}$ exceeds the one for YbRu$_2$Si$_2$ ($b=0.25$) \cite{Trovarelli2000}. This implies perhaps a different character of the interplay between the critical fluctuations and fermionic degrees of freedom. 

The challenge in obtaining the scaling shown in Fig.~\ref{fig:Scaling09} was to determine the normal-state zero-field $\gamma(T,0)$ at low temperatures since this sample becomes superconducting for temperatures below $\sim$ 2 K. We overcame this problem by determining the $T$ dependence of the metallic $\gamma(T,5$ T) down to 0.5 K. We obtained a power-law $T$ dependence of the form $\gamma(T,5$ T$)=0.52/T^{0.48}$ by fitting these data. (see blue line in inset to Fig.~\ref{fig:Scaling09}). These data are in the NFL regime over this temperature range, with a crossover to the FL regime at $T \approx 0.4$ K (see Fig.~\ref{fig:PhaseDiagram09}). This procedure is supported by the fact that the normal-state $\gamma(T,H)$, exposed at low temperature by the application of a magnetic field, follows a power law behavior, clearly visible for fields up to 8 T and temperatures as low as 0.5 K, as shown in the inset to Fig.~\ref{fig:HighHC09}. In addition, we confirmed that the normal-state $\gamma(T,0)$ of Ce$_{0.91}$Yb$_{0.09}$CoIn$_{5}$ diverges as $\gamma(T,0)\approx0.52/T^{0.48}$, not as $-\log T$, as $T\rightarrow 0$ by showing that the normal-state and superconducting entropies are equal at $T_c$ (see discussion and figure in Supplementary Materials section).

This scaling analysis, together with the phase diagram of Fig.~\ref{fig:PhaseDiagram09}, generated based on the FL to NFL crossover in $\gamma(T)$ and $R(T)$, serve as evidence that there is a second order QPT with $H_{\textrm{QCP}}=0$ for the $x_c=0.09$ Ce$_{1-x}$Yb$_{x}$CoIn$_5$ alloy. Hence, Ce$_{0.91}$Yb$_{0.09}$CoIn$_5$  is, indeed, a zero field quantum critical metal at ambient pressure. 
We have shown previously that $H_{QCP}$ of the Ce$_{1-x}$Yb$_{x}$CoIn$_5$ system is suppressed from about 4.1 T for CeCoIn$_5$ to close to zero for $x = 0.07$ \cite{Hu2013}. This is consistent with the present finding that the higher doping of $x=0.09$ ($x_n=0.25$) is the critical doping. In addition, changes in the Fermi surface topology of Ce$_{1-x}$Yb$_{x}$CoIn$_5$ have been revealed by de Haas-van Alphen studies, which found the disappearance of the intermediately heavy $\alpha$ sheet for $0.1\leq x_n \leq 0.2$ \cite{Polyakov2012}. This result correlates with a study of the electronic structure of Ce$_{1-x}$Yb$_{x}$CoIn$_{5}$, which has shown that the Yb valence for $x_n \leq 0.2$ increases rapidly  from Yb$^{2.3+}$ toward Yb$^{3+}$ with decreasing $x$ \cite{Dudy2013}. Also, recent penetration depth measurements on Ce$_{1-x}$Yb$_{x}$CoIn$_{5}$ have reported that the superconducting gap changes its character around ${x \approx 0.037}$ ($x_n=0.2$) from nodal to nodeless  with increasing Yb doping \cite{Kim2015}. 
Hence, all these findings indicate that the following features  are present at $x_c\approx 0.09$ ($x_n=0.25$): Fermi surface reconstruction,  Yb valence transition from Yb$^{3+}$ to Yb$^{2.3+}$ (the Ce valence is 3+ for all $x$ values), transition from nodal to nodeless superconductivity, and suppression of $H_{QCP}$ towards 0 T. We note that $T_c(x)$ decreases linearly with $x$ \cite{Hu2013}, without any features near $x_c=0.09$. 

Based on these results, one is tempted to speculate that the nodal gap structure and unconventional superconductivity in Ce$_{1-x}$Yb$_{x}$CoIn$_{5}$ can be attributed to the presence of AFM critical spin fluctuations  near a QCP for $x<x_c=0.09$. We note that in many cases critical spin fluctuations lead to the formation of a nodal gap~\cite{Hashimoto2013}. With increasing Yb doping beyond $x_c$, this system displays conventional SC in which the emergence of SC and onset of many-body coherence in the Kondo lattice have same physical origin: hybridization between conduction and localized Ce $4f$-electron states \cite{Hu2013}.  We note that the presence of Yb as the substitution for Ce provides a unique scenario in which quantum criticality is observed for Yb doping for which Yb exhibits a magnetic valence. In this sense, the magnetic valence of Yb might have a role in preserving the long-range order of the diluted Ce lattice, and, thereby, facilitating the magnetic order and the development of a quantum phase transition in this system. At the same time, the robustness of unconventional superconductivity with respect to disorder points out towards the multiband nature of superconductivity: intraband disorder scattering is dominant, while pairing
involves several bands and therefore remains largely immune to disorder.

Our analysis of the specific heat data for $H=0$ also reveals a discontinuous change in slope of $C/T$ as a function of $T$ at $T_{c3}\approx 7$ K (right inset of Fig.~\ref{fig:Scaling09}). This means that the third derivative of the free energy with respect to temperature is discontinuous and, therefore, the system undergoes a \emph{third-order phase transition} at $T=T_{c3}$. We leave for future studies the possible origin of this transition and whether it is governed by the underlying quantum 
critical fluctuations.

\paragraph{Summary.} To conclude, we performed specific heat, electrical resistance, and  magnetoresistivity measurements on single crystals  of the heavy-fermion superconducting alloy Ce$_{0.91}$Yb$_{0.09}$CoIn$_5$ and have shown that this material is quantum critical, i.e., it has an antiferromagnetic QCP in zero magnetic field and at ambient pressure. Hence, the physical properties of this material in the normal state at low temperatures are controlled by the presence of this QCP. The existence of this QCP is confirmed by the scaling analysis of the specific heat data. The AFM nature of this QCP is suggested by the excellent fits of both heat capacity and resistance data measured in different magnetic fields with the spin fluctuation theory \cite{Takimoto1995}.  Our findings,  along with other recent reports on this  system, suggest that the origin of the superconducting pairing is different in samples with low and high Yb doping: The presence of AFM fluctuations are most likely the reason for the nodal gap at lower doping, while the fact that $x_c=0.09$ for Ce$_{1-x}$Yb$_{x}$CoIn$_{5}$, hence there are no AFM fluctuations for $x \geq 0.09$, implies that a conventional pairing mechanism gives the nodeless characteristics of the superconducting gap.  

This work was supported by the National Science Foundation grants DMR-1505826 and DMR-1506547 at KSU and by the
US Department of Energy, Office of Basic Energy Sciences, Division of Materials Sciences and Engineering, under Grant
Nos. DE-SC0016481 at KSU and DE-FG02-04ER46105 at UCSD. MD also acknowledges the hospitality of Kirchhoff Institute of Physics (University of Heidelberg), where part of this work has been completed. 

\bibliography{References}

\newpage
\widetext
\section{Supplementary Materials}
\section{Magnetoresistivity}
The $H$ dependence of MR is non-monotonic in a Kondo lattice material: a positive MR at low fields is followed by a peak in MR at $H_{max}^{MR}$ and negative and quadratic (in $H$) MR at high fields. It is natural to think that the low-field positive MR is a result of coherence developing in the system.  
The temperature dependence of $H_{max}^{MR}$ in Kondo lattice systems can be used to extract information regarding the nature of the ground state of Ce$_{1-x}$Yb$_{x}$CoIn$_{5}$ alloys \cite{Hu2013,Singh2014,Singh2015}. 
Specifically, $H_{max}^{MR}$ increases with decreasing $T$ in conventional Kondo lattice systems, while, $H_{max}^{MR}$ decreases with decreasing $T$ in unconventional Kondo lattice systems that exhibit quantum critical properties. In this latter case, the extrapolation of $H_{max}^{MR}$  to zero temperatures  gives $H_{\textrm{QCP}}$  \cite{Hu2013}. 

\begin{figure}
\centering
\includegraphics[width=0.4\linewidth]{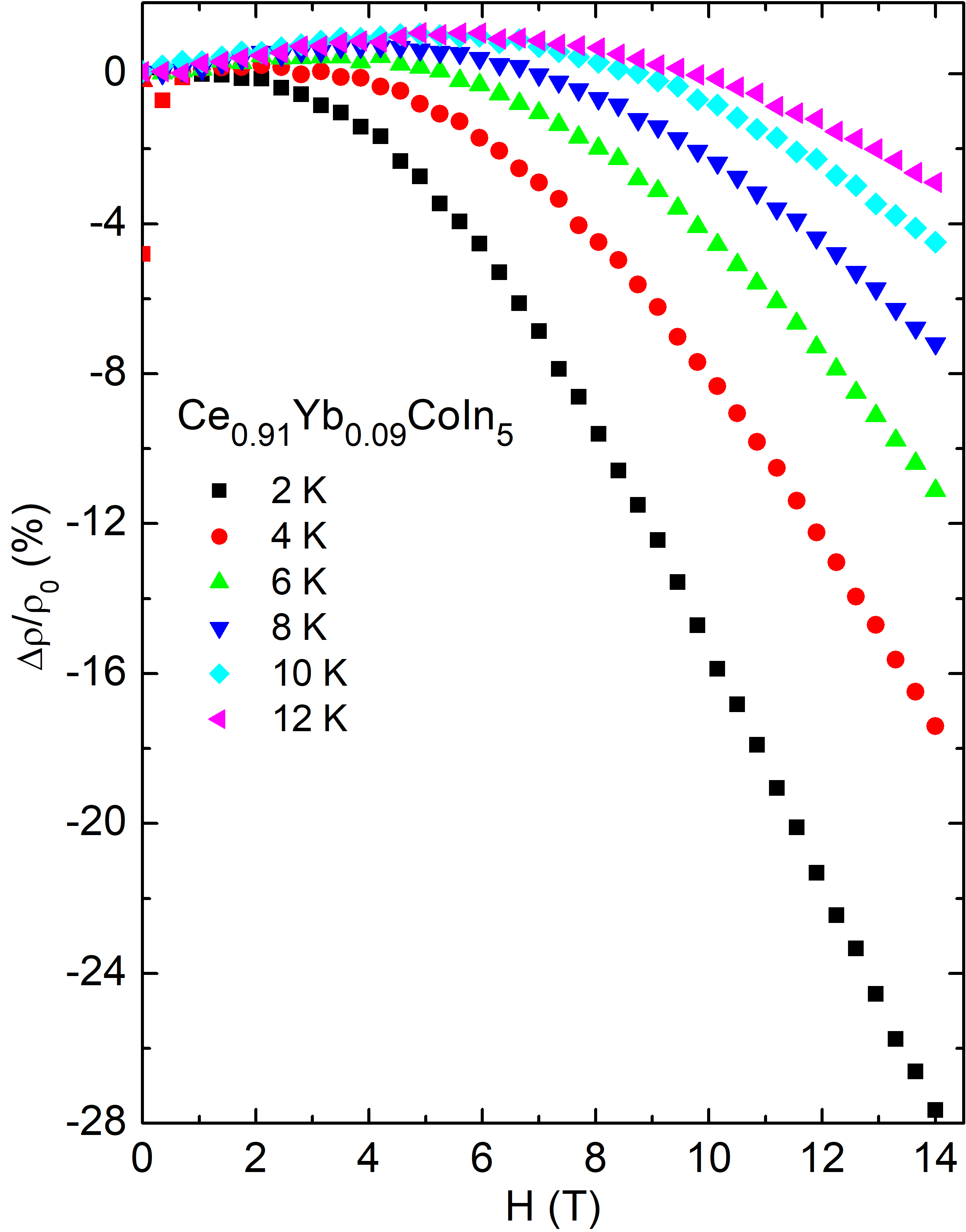}
\caption{Magnetoresistivity $\Delta \rho/ \rho_0$ of Ce$_{0.91}$Yb$_{0.09}$CoIn$_{5}$ plotted as the function of magnetic field $H$, measured at different temperatures.} 
\label{Supp-Fig-1}
\end{figure}

Figure~\ref{Supp-Fig-1} shows the transverse magnetoresistivity (MR) of Ce$_{0.91}$Yb$_{0.09}$CoIn$_{5}$. This reveals a low-field positive contribution to MR, in addition to the high-field quadratic and negative in field MR \cite{Hu2013}. Also, notice that the position ($H_{max}^{MR}$) of the peak in MR  decreases with decreasing temperature for $T\leq 12$ K. As mentioned above, this unconventional behavior of magnetoresistivity in the Ce$_{1-x}$Yb$_{x}$CoIn$_{5}$ Kondo lattice system is indicative of a significant contribution to scattering coming from quantum fluctuations. Therefore, the analysis of transverse MR data of Ce$_{0.91}$Yb$_{0.09}$CoIn$_{5}$  reveals the unconventional $H_{max}^{MR}$ dependence and confirms the quantum critical nature of these material: a linear extrapolation of $H_{max}^{MR}(T)$ (blue triangles in Fig.~2 of main text), which separate the regions with positive ($d\rho/dH>0$) and negative ($d\rho/dH<0$) magnetoresistivity within the NFL regime, to zero temperature again results in $H_{\textrm{QCP}} = 0$ T, showing that $x_c=0.09$ for the Ce$_{1-x}$Yb$_{x}$CoIn$_{5}$ system.

\section{Theoretical analysis of the experimental data}
In this Section we provide the details for the theoretical fitting of the specific heat and resistance data. Our starting point is the phenomenological theory
by Moriya and Takimoto \cite{Moriya1995}, which assumes that the anomalous thermodynamic and transport properties of heavy electron systems are governed by the system's proximity to a quantum phase transition between the antiferromagnetic and paramagnetic states. 
In particular, it assumes that the dynamical magnetic susceptibility due to heavy-electrons is purely local:
\begin{equation}\label{chiLocal}
\chi_L(\omega)=\frac{\chi_{L}}{1-{i\omega}/{\Gamma_L}}.
\end{equation}

Another important quantity is the staggered inverse magnetic susceptibility 
\begin{equation}
y=\frac{1}{2T_A\chi_Q},
\end{equation}
where $T_A\approx(J_{\vec Q}-J_{{\vec Q}+{\vec q}_B})/2$ and $J_q$ is the antiferromagnetic exchange coupling between the localized magnetic moments, ${\vec Q}$ is the antiferromagnetic wave vector, and $|{\vec q}_B|$ is the zone-boundary vector.    
The temperature dependence of the function $y$ is given by \cite{Moriya1995}:
\begin{equation}\label{yt}
y=y_0+\frac{3}{2}\int\limits_0^1x^2\left[\log u-\frac{1}{2u}-\psi(u)\right]dx, \quad u=\frac{y+x^2}{t},
\end{equation}
where $y_0=1/2T_A\chi_{Q}(0)$ effectively measures the distance to the quantum critical point (QCP) so that $y_0=0$ at QCP,
$t=T/T_0$ is the reduced temperature, $T_0=T_A\Gamma_L\chi_{L0}/\pi$, where $\chi_{L0}$ is the local magnetic susceptibility at $T=0$, and $\psi(u)$ is the digamma function defined as the logarithmic derivative of the gamma function, i.e., $\psi(z)=\Gamma'(z)/\Gamma(z)$. 
\begin{figure}[h]
\centering
\includegraphics[width=12cm,height=9cm,angle=0]{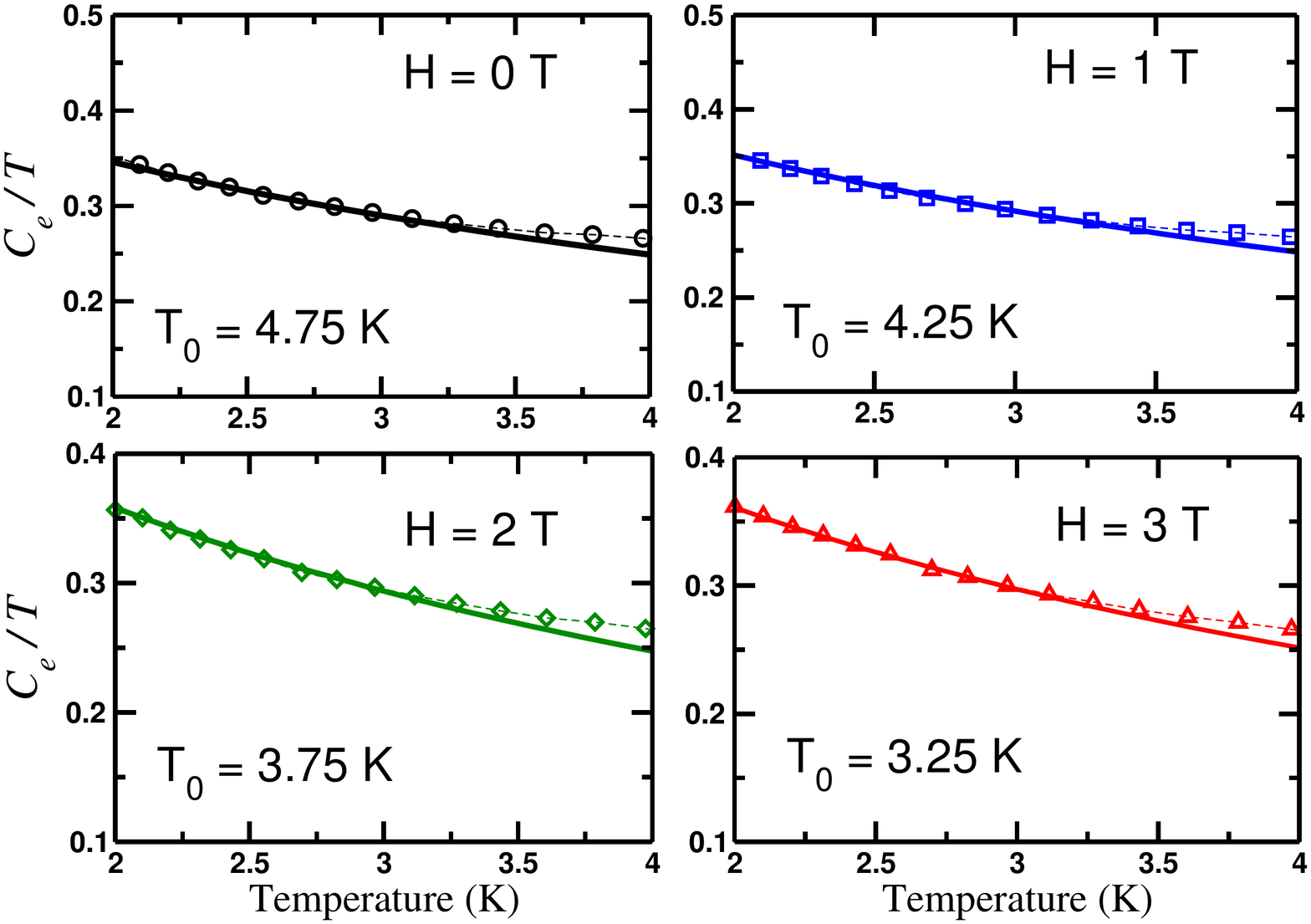}
\caption{\small(Color online) Theoretical fits to the heat capacity data. In all four plots $k_BT_0(H=0)\nu_F\approx 5$. In addition, the values of the parameter $y_0$ are: $y_0(H= 0 \textrm{T})=0.0$, $y_0(H= 1\textrm{T})=0.01$, 
$y_0(H= 2 \textrm{T})=0.03$, and $y_0(H= 3 \textrm{T})=0.04$.}
\label{Supp-Fig-2}
\end{figure}
\begin{figure}[h]
\centering
\includegraphics[width=12cm,height=9cm,angle=0]{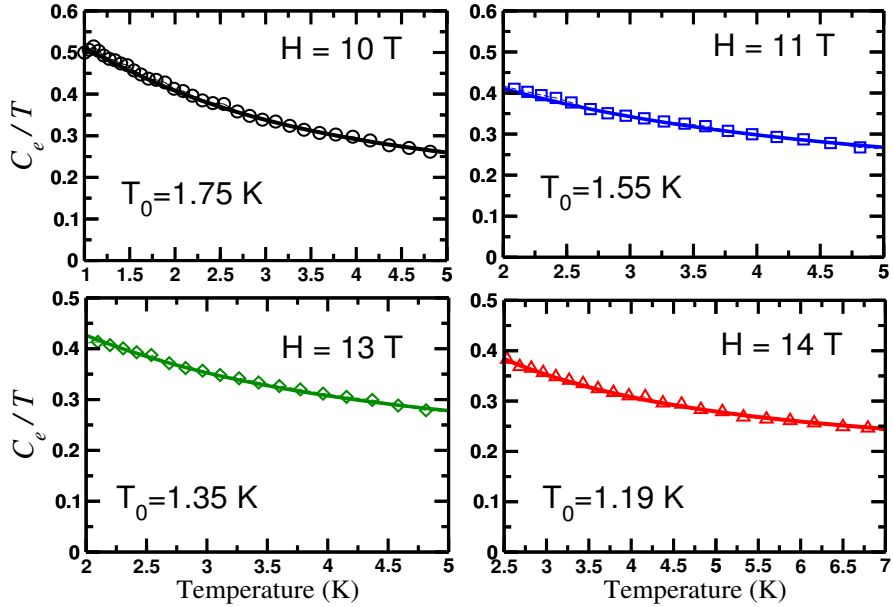}
\caption{\small(Color online) Theoretical fits to the heat capacity data. In all four plots $k_BT_0(H=0)\nu_F\approx 5$. In addition, the values of parameter $y_0$ are:  $y_0(H= 10 \textrm{T})=0.15$, $y_0(H= 11 \textrm{T})=0.18$, 
$y_0(H= 13 \textrm{T})=0.21$, and $y_0(H= 14 \textrm{T})=0.25$.}
\label{Supp-Fig-3}
\end{figure}
\subsection{Heat capacity}
Having defined all necessary parameters, we use the following expression for the electronic specific heat to fit our data:
\begin{equation}\label{CmT}
\begin{split}
C_{\textrm{e}}=9\nu_F\int\limits_0^1x^2&\left\{\left[u^2-2u\frac{dy}{dt}+\left(\frac{dy}{dt}\right)^2\right]\left[-\frac{1}{u}-\frac{1}{2u^2}+\psi'(u)\right]\right.\\&-\left. t\frac{d^2y}{dt^2}\left[\log u-\frac{1}{2u}-\psi(u)\right]\right\}dx,
\end{split}
\end{equation}
where $\nu_F$ is the electronic density of states at the Fermi level. We note that we used equations (\ref{yt}) and (\ref{CmT}) to fit our data using $\nu_F$, $T_0$ and $y_0$ as fitting parameters. The results are shown on Fig. \ref{Supp-Fig-2}  in the low-field range ($0$ T $\leq H\leq 3$ T) and Fig. \ref{Supp-Fig-3} for relatively high fields ($10$ T $\leq H\leq 14$ T). The following comments are in order:   (i) fairly good fits for the heat capacity at low fields definitely hint that the underlying QCP is of magnetic origin and, in addition, to the fact that the magnetic field is effective in suppressing the quantum critical fluctuations; (ii) while valence fluctuations of the Yb ions definitely contribute to $C_{\textrm{e}}/T$, the leading contribution is governed by the underlying magnetic fluctuations of the under-screened Ce moments; (iii) the relatively weak effect of Yb alloying on superconductivity may be due to the combination of two factors: intrinsic quantum critical fluctuations around $H_{\textrm{QCP}}\approx 0$, which likely lead to pairing and the multiband nature of superconductivity in which the disorder scattering is different for different Fermi surfaces. 

\subsection{Resistivity}
\begin{figure}
\centering
\includegraphics[width=0.5\linewidth]{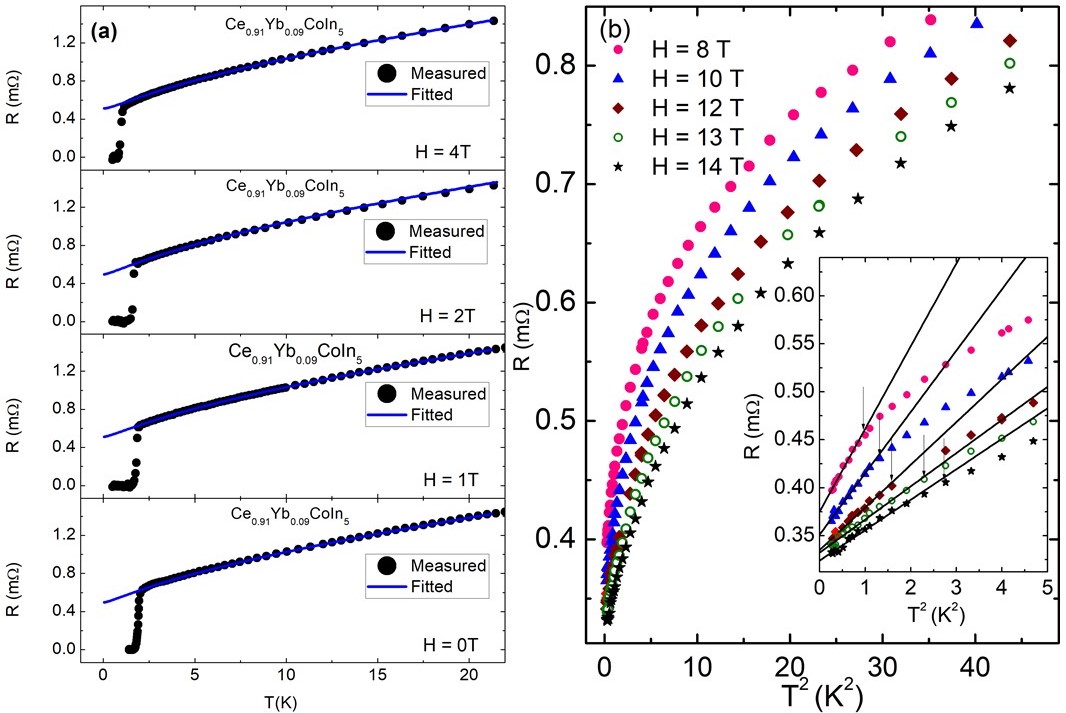}
\caption{\small(Color online) Plots of resistance $R$ as a function of temperature $T$ (left panel) and $T^2$ (right panel) measured at various transverse applied magnetic field $H$ for Ce$_{0.91}$Yb$_{0.09}$CoIn$_{5}$ single crystal. The fits to the data of the left panel have been done using the 
phenomenological model of antiferromagnetic quantum criticality \cite{Moriya1995} . The solid lines on the inset of the right panel are fits to the low $T$ data and the arrows indicate the point where the data deviate from the $T^2$ dependence fits.}
\label{Supp-Fig-4}
\end{figure}

We use the theory of Moriya and Takimoto \cite{Moriya1995}, which incorporates the effect of spin fluctuations on scattering in the vicinity of a second order quantum phase transition, to fit the low-temperature resistance data. In doing so, we assume that the main inelastic contribution to the scattering originates from the Zeeman coupling of magnetic field to the spin-degrees of freedom and, consequently, ignore the orbital quantization effects emerging at high magnetic fields. The corresponding expression for resistivity we employ reads
\begin{equation}\label{rhoT}
\rho(t)=\rho_a\int\limits_0^1 dx\left(-1-\frac{1}{2u}+u\frac{d\psi}{du}\right)dx+\rho_b,
\end{equation}
where $\rho_{a}$ along with $y_0$ are the field-dependent fitting parameters and $\rho_b=0.5$ (m$\Omega\cdot\textrm{cm}$). We employ $\rho_b$ to take into account the effects of scattering induced by Yb substitution.  
We note that we fit the resistance data measured in the low-field range ($0\leq H\leq 4$ T) since at higher fields the effect of magnetic fluctuations on resistivity is suppressed and the scattering due to electron-phonon coupling as well as electron-electron interactions dominates the temperature dependence of resistivity. As a consequence, fits of the resistance data measured in high magnetic fields using expression (\ref{rhoT}) are not expected to work well and, as we have verified, they do not. Indeed, the right panel of Fig.~\ref{Supp-Fig-4}  shows that the application of high magnetic fields suppresses the spin fluctuations and gives rise to a $T^2$ dependence in $R$ at low temperatures, revealing a low temperature  FL regime.

The results of our fits are shown on the left panels of Fig. 4. We were able to obtain exceptionally good fits with only one fitting parameter $y_0$, while $\rho_a$ remains very weakly dependent on the value of the magnetic field. In fact it only slightly decreases with increasing $H$ from $\rho_a\approx 0.372$ (m$\Omega\cdot\textrm{cm}$) at $H=0$ T to $\rho_a\approx 0.359$ (m$\Omega\cdot\textrm{cm}$) for $H=4$ T. Based on these results, we conclude that the system's proximity to the zero-field quantum critical point is a very likely source for the anomalous thermodynamic and transport properties. 

A further confirmation of a crossover from FL to NFL behavior in Ce$_{0.91}$Yb$_{0.09}$CoIn$_{5}$ is also given by the electrical resistance data. The right panel to Fig.~\ref{Supp-Fig-4} and its inset reveal a FL to NFL crossover.  We denote $T_{\textrm{FL}}^R$ [see arrows on right panel of Fig.~\ref{Supp-Fig-4}] the temperatures at which the transition from the low temperature FL to the higher temperature NFL behavior takes place. 

\section{Entropy}

\begin{figure}
\centering
\includegraphics[width=0.5\linewidth]{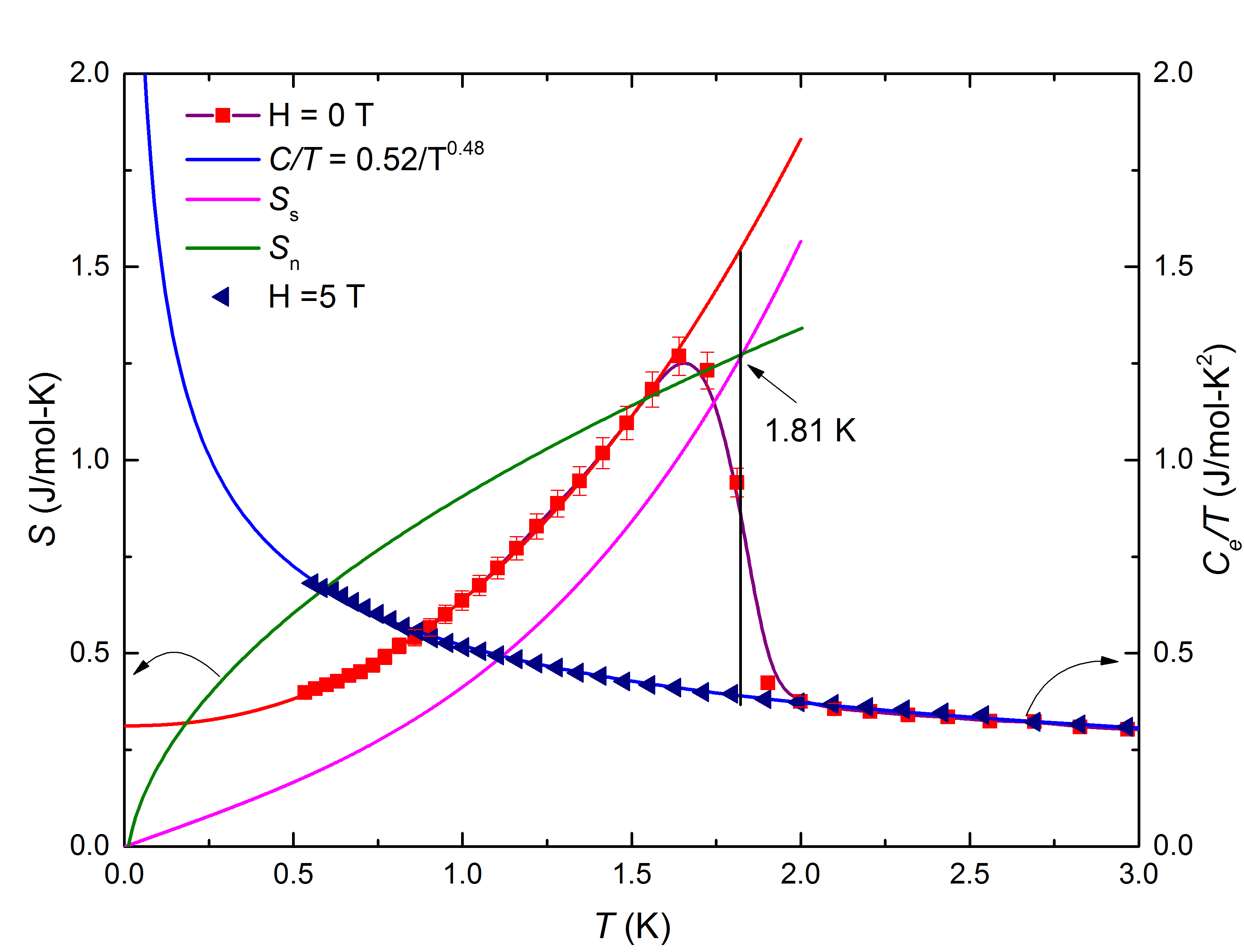}
\caption{\small(Color online) Plots of entropy $S$ (left-axis) and $C_e/T$ (right-axis) as a function of temperature $T$. The red and blue curves are the fits to the 0 T superconducting data and 5 T normal state data respectively. The green and magenta curves are the temperature dependence of entropy from normal and superconducting states respectively.}
\label{Supp-Fig-5}
\end{figure}

 A further confirmation that the normal-state $\gamma(T,0)$ of Ce$_{0.91}$Yb$_{0.09}$CoIn$_{5}$ diverges as $\gamma(T,0)\approx0.52/T^{0.48}$, not as $-\log T$, is given by the calculation of the low temperature entropy $S$. To extract the superconducting entropy from our low-$T$ heat capacity data, we first estimated the $C_e/T$ vs. $T$ down to 0 K  by fitting the  superconducting 0 T data over the range $0.5 \leq T \leq 1.6$ K (red curve in Fig.  \ref{Supp-Fig-5}) and extrapolating this fit to 0 K. To extract the normal-state entropy, we fitted the normal-state heat capacity data measured in 5 T  over the range $0.5 \leq T \leq 7$ K with $C_e/T$ = $0.52/T^{0.48}$ (blue curve in Fig. ~\ref{Supp-Fig-5}) and  extrapolated this fit to 0 K. Our use of the 5 T data to calculate the normal-state entropy in zero field is justified by the fact that the 0 T and 5 T specific heat data overlap in the normal state, however the normal-state $T$ range is very limited in zero field since the data reveal superconductivity below 2.3 K .
We then calculated the entropy as
\begin{equation}
\begin{split}
{S}= \int_0^{T}\frac{C_e}{T}{d}T,
\end{split}  
\end{equation} 
where $C_{e}\equiv C-C_{ph}$. Figure~\ref{Supp-Fig-5} shows the plots of the normal-state of $S_n$ (green curve) and superconducting $S_s$ (pink curve) entropies. Notice that the two entropies are equal to 1.24 J/mol-K (the two curves cross) at 1.81 K. This temperature is the superconducting transition temperature as determined using the isentropic method at $T_c$.
We note that a $-\log T$ divergence of $C_e/T$ does not balance the normal-state and superconducting entropies at the superconducting transition temperature. In addition, we observe that $\lim\limits_{T\to 0}(C(T)/T)$ remains finite, which is likely due to the combination of two main factors: the nodal structure of the superconducting gap function $\Delta({\mathbf k})$ and disorder-induced single particle states due to ytterbium substitution.

\end{document}